# Effect of NaCl on *Pseudomonas* biofilm viscosity by continuous, non-intrusive microfluidic-based approach

F. Paquet-Mercier,[a] Parvinzadeh Gashti,[a] M. J. Bellavance,[a] S.M. Taghavi,[b] J. Greener[a,*]

A method combining video imaging in parallel microchannels with a semi-empirical mathematical model provides non-intrusive, high-throughput measurements of time-varying biofilm viscosity. The approach is demonstrated for early growth *Pseudomonas sp.* biofilms exposed to constant flow streams of nutrient solutions with different ionic strengths. The ability to measure viscosities at early growth stages, without inducing a shear-thickening response, enabled measurements that are among the lowest reported to date. In addition, good time resolution enabled the detection of a rapid thickening phase, which occurred at different times after the exponential growth phase finished, depending on the ionic strength. The technique opens the way for a combinatorial approach to beter understand the complex dynamical response of biofilm mechanical properties under well-controlled physical, chemical and biological growth conditions and time-limited perturbations.

## Introduction

Bacterial biofilms consist of bacteria surrounded by an adaptive extra-cellular polymeric substance (EPS) with complex physio-chemical properties. Believed to be among the first multi-cellular life-forms on Earth, biofilms are well-represented in nature, ranging from slimy coatings on rocks in rivers to oral-dwelling variants. Biofilms are also interesting from the perspective of new bio-technical applications, because they are essentially natural, self-repairing catalytic materials, which operate at ambient conditions.[1,2] The biofilm EPS is integral to its survival by forming a protective barrier, which regulates penetration of foreign chemical species, release of metabolic byproducts and planktonic cells and more.[3,4] Their complex mechanical properties have been a point of focus for the last 25 years.[5,6] However, despite the recognition that biofilms have both viscous and elastic properties,[7,8] most work has focused on the latter. Viscous behaviour can be important for surface spreading and contamination[9] and in forming elongated structures such as streamers.[10,11] It has also been shown that many bacterial biofilms demonstrate a transition from elastic to viscous after 18 minutes following application of elevated shear stress. Since this relaxation time is short compared to the duration of applied shear in typical flow-cell measurements, viscous behaviour should be important. Valuable measurements of biofilm viscous properties have been obtained using standard techniques such as parallel plate rheometry,[12] cone and plate rheometry,[13] force transducers such as AFM[14,15] and generation of stress-strain curves in flow cells.[8,16,17] However, all of these techniques impose external forces that can bias results due to the well-known non-Newtonian properties of biofilms.[6,8] In addition, most techniques do not replicate standard flow cell or natural growth environments. Measurements by pendant drop tensiometry have been conducted, but had relevance to biofilms at liquid/gas interfaces.[18]

Despite the computational resources required, microrheology offers an exciting emerging method for combining optical imaging with localized, *in situ* rheological measurements for biofilms.[9,19] However, the use of two-particle correlation method to ensure that bulk properties are measured has not yet been demonstrated.[6,20]

In this work we present a method for measuring the viscosity of a growing biofilm, which combines microfluidics and standard optical microscopy with a straight-forward fluid dynamic model. The model requires two measurements: the downstream velocity of moving biofilm segments and their height. The latter is estimated from calibrated measurements of optical density, with validation being provided by confocal laser scanning microscopy (CLSM). The technique is used for a high-throughput, proof-of-principle study of the effect of ionic strength on the time-varying viscosity in young *Pseudomonas sp.* biofilms.

## Experimental section

### Materials

Photoresists (SU8 3000, Microchem, USA), sealing glass (Fisher Scientific, USA), tubing (Hamilton, Canada), 60 mL polypropylene syringes (BD, USA), tryptone (Sigma-Aldrich, USA), yeast extract (Sigma-Aldrich, USA), sodium chloride (Caledon Laboratories Ltd, Canada). Sterile distilled water for making media, and **the** bacteria were *Pseudomonas sp.*, strain CT07 with a green fluorescent protein (GFP) expressing tag.

### Microfluidic device

A multi-channel microfluidic flow cell was fabricated by casting polydimethyl siloxaine (PDMS) against a template mould, which consisted of micro pattered photoresist features (SU8-3000) on a silicon wafer support substrate (FlowJEM, Canada). The resulting patterned PDMS was then sealed using a microscope slide (130 μm) after plasma activation (PCD-001 Harrick Plasma, Ithaca, NY, USA). The height, width and length of the microchannel for optical microscopy were $h_c$=400 μm, $w_c$=3000 μm and $l_c$=25 mm, respectively. The microchannel for CLSM had the same cross-section but had a length of 2.5 mm. Liquid was injected by a six-syringe load pump (NE-1600, New Era, USA). See Supporting Information for more information.

[a.] Département de Chimie, Université Laval, Québec, QC, G1V 0A6, Canada.
[b.] Département de Génie Chimique, Université Laval, Québec, QC, G1V 0A6, Canada
*Corresponding author

## ARTICLE

### Inoculation and growth process

Colonies of *Pseudomonas sp.* were cultured at 30 °C for 3 days (name of incubator) in a LB nutrient-containing Petri dish. Pre-cultured bacteria were used as inoculum for seeding. The bacteria colonies were scraped from the surface of LB plate and transferred to a tube containing 3 mL of a modified LB medium, which consisted of 0.1 wt% tryptone, 0.05 wt% yeast extract and NaCl concentration of 0.1 wt%. The tube was incubated for 18 h on an orbital shaker (mini-shaker, VWR International) at 30°C, at 300 rpm. The tube connectors (SciPro, Canada) and microfluidic channels were disinfected by flowing 70 wt% ethanol for 2 h and then sterile distilled water (autoclaved) for 1h with flow rate of 2 mL·h$^{-1}$. Inoculation was achieved by dividing the same pre-culture into four 1 mL syringes and flowing into the channels for 30 min at 0.2 mL·h$^{-1}$. After inoculation, the channels were exposed to a fresh modified LB nutrient with flow rate $Q_{nutr}$ = 0.2 mL·h$^{-1}$. All modified LB solutions were comprised of 0.1 wt% tryptone and 0.05 wt% yeast extract, and NaCl concentrations of 0 wt%, 0.05 wt% (8.5 mM), 0.1 wt% (17 mM) and 0.2 wt% (34 mM).

### Characterization

Acquisition of optical micrographs was accomplished with an inverted microscope (IX73, Olympus, Canada) with 2x objective and an automatic stage (MS2, Applied Scientific Instruments MS2, USA). Both image acquisition and stage positioning were controlled by a customized macro within the imaging software environment (Image ProPlus 7.0, Media Cybernetics, USA). Image analysis was conducted using the Fiji bundle for ImageJ[21] and the particle tracking algorithm MTrackJ.[22] Optical density (OD) measurements were acquired from pixel intensities after a calibration process using target with known OD values. Acquisition of 3D confocal stacks was achieved with a confocal laser scanning microscope (FV1200, Olympus, Japan) with a 10x objective and a PMT detector. The GFP was excited at 488 nm and emission was detected in between 500-600 nm. Image stacks were analyzed using the Fiji bundle for ImageJ. Height measurement from CLSM 3D confocal stacks were achieved by determining the first and last slice where the biofilm was visible (see Supporting Information for more information).

### Computer model

The model that relates displacement of moving biofilms with their viscosities was developed previously.[11] A spreadsheet was developed with built in formulas to efficiently calculate time-varying viscosity based on experimentally measured parameters in this work (Supporting Information).

## Results

### Time-lapse videos

After inoculation, the flowing LB nutrient solution produced a shear stress against all walls of the microchannel and the biofilms growing on them. This was the force responsible for the displacement of biofilm segments in time. Fresh nutrient was consumed by the biofilms and contributed to accumulation of biomass and a related increase in optical density (OD). Both effects were captured by microscope imagery every 30 mins for between 60-80 hours to create time lapse videos (Figure 1). Each 8-bit, 5-megapixel image was acquired using the same exposure time, illumination intensity and gain. As an example, Figure 1b shows an example of five sequential frames from a time-lapse video.

Due to the low-magnification used in these studies (2x objective), the biofilm could only be observed after the aggregation of sufficient quantities of biomaterial resulted in segments with relatively high OD. In addition, tracking accuracy was limited due to pixel sizes of 3 μm x 3 μm. Thus, with imaging every hour, the lowest measurable velocity could be 3 μm·h$^{-1}$. In practice this limit was even higher because biofilm segments could not be tracked with single-pixel accuracy. In addition, the evolution of OD of tracked biofilm segments was also monitored in time (Supporting Information).

### Biofilm displacement and velocity

The displacement of tracked biofilm segments and their changing OD were measured by particle tracking. In the early times after inoculation, growth had a tendency to expand the clusters in all directions. We measured the expansion in both the downstream and upstream directions by tracking the displacement of the clusters' leading edges, $d_{down}$ and $d_{up}$, respectively, during this period (Figure 2). The net displacement downstream, $d_{net}$, was calculated from:

$$d_{net} = d_{down} - d_{up} \quad (1a)$$

$$d_{net} = d_{down} \quad (1b)$$

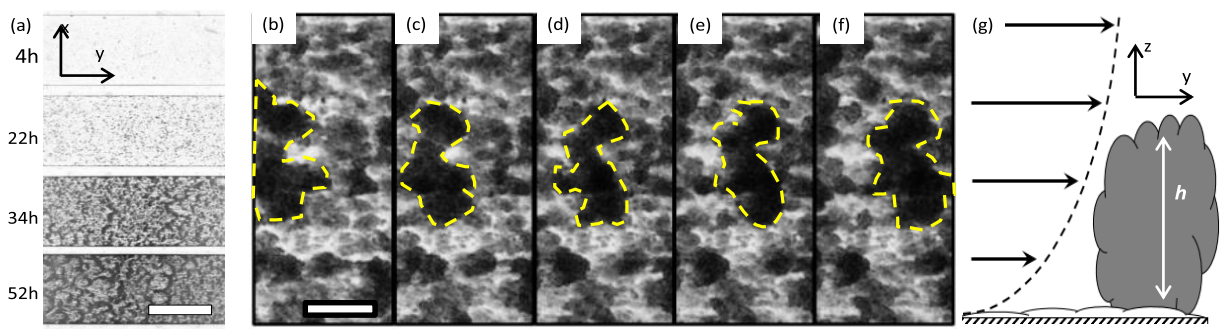

**Figure 1.** (a) Images in the x,y plane of growing biofilm acquired at indicated times after inoculation. Modified LB nutrient flow (0.1 wt% NaCl) was in the y-direction. Scale bar is 1500 μm. Time-series optical micrograph from the same data set as (a) at 33 h-37 h after inoculation (b-f). Scale bar in (b) is 150 μm and is representative for (b-f). (g) Schematic showing a cross-section (z,y plane) of the moving feature shown in (b-f) with height h and a superimposed parabolic liquid flow velocity profile (dashed) that is normally in place in the absence of obstructions.

ARTICLE

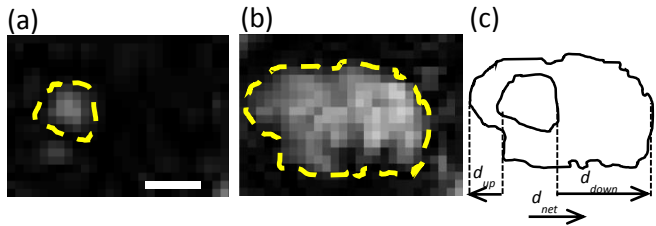

**Figure 2**. The growth of a representative biofilm colony between 17h (a) and 24h (b) in the same field of view under modified LB medium with 0.1 wt% NaCl. Its boundary is highlighted with a dashed line for eye guidance. (c) The colony boundary lines are superimposed on one-another revealing the differences in $d_{up}$ and $d_{down}$. The $d_{net}$ approximates the net displacement of the colony's centre points. Scale bar in (a) is 25 μm, representative for all sub-figures.

At early growth times (t<10 h after lag phase), $d_{up}$ was approximately 50 wt% of $d_{down}$. During later times, $d_{up}$ was negligible compared to $d_{down}$ and did not significantly contribute to $d_{net}$ so Eq. (1b) to calculate $d_{net}$.

This was repeated for each of the tracked biofilm segments and the average value of $d_{net}$ was calculated as

$$\tilde{v}_{net} = \frac{\left(\sum_i^n \frac{d_{net,i}}{n}\right)}{\Delta t} \quad (2)$$

Where the summation on the right determined the average net displacement for $n$ tracked biofilm segments.

Similar movement, referred to as "peel and move" was noted before, describing a sliding motion that maintained biofilm contact with the surface, but changed its attachment surface position.[23] Instead, we noted here that the biofilm remained stuck in place while its leading edge streaked down the channel (similar to Figure 2, but more pronounced at longer times). We believe a "grow and flow" movement justifies the use of a viscous flow model of the moving biofilms in our case. Therefore, the viscosity measured here is likely localized near the attachment surface. Due to inconsistencies in the flow velocity profile, the tracked velocity near vertical side-walls differed from those in the middle of the channel (Supporting Information). Therefore, for the experiments reported here, velocities were measured at a minimum distance of 150 μm from either vertical side-wall.

**Estimating height of tracked biofilm segments**

The second input to the mathematical model is the average height (*h*) of the tracked biofilm segments. In order to be as generalizable as possible, the method should be implemented by regular time-lapse imaging using a regular transmission microscope with a fixed focal position. However, the drawback is that biofilm heights are not directly measured. To overcome this, we estimated *h* based on the measured biofilm OD, which is directly proportional to the total biomass.[24] This was accomplished using a two-point calibration method. First, OD was measured at the beginning of the experiment, before any biofilm had accumulated ($h_{min}$ = 0 μm). Then, OD values were measured in the channel corners (beside the vertical side-wall) near the end of the experiment, when we were sure that biofilm had formed a continuous sheet from the bottom of the microchannel to the top with a maximum height ($h_{max}$ = 400 μm). As an example, we transform the OD for the biofilm grown under concentrations for 0.1 wt% NaCl into height measurements using the following calibration curve and plot the results in Figure 3:

$$h = 153 \cdot OD - 0.6 \ (\mu m) \quad (3)$$

It is important to note that calibration equations in the form of eqn. (3) considers that all increases to OD are due to increases from *h*, which ignores contributions from densification. It is known that biomass at the attachment surface, is generally the most dense, usually due to polysaccharide accumulation over time, whereas the opposite is true for the top portion of the biofilm, which is younger.[25,26] Therefore, on the balance, our simplification may not be too unreasonable, especially over the time course of just a few days.[27] As a validation step, we present the results from CLSM conducted during a separate experiment superimposed in Figure 3. The experimental window was limited to the first 30 hours until *h* stopped increasing. It is clear that the two curves show similar behaviour in terms of both final $h_{max}$ as well as the rate of change in *h* during the rapid growth phase. Differences at in *h* at early times following the end of the lag phase likely resulted from insufficient z-direction resolution in CLSM compared to young thin biofilm segments. This demonstrating an unexpected benefit of using calibrated optical density over direct confocal measurements.

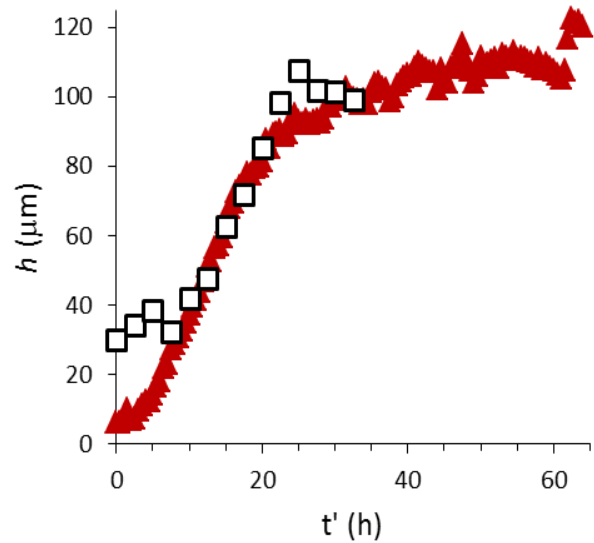

**Figure 3.** Average height measurements from calibrated OD (red triangles) and CLSM (white squares) for growing *Pseudomonas sp*. Measurements were acquired from different pre-culture bacterial samples, but were grown under the same experimental conditions (Q=0.2 mL·h⁻¹, [NaCl]=0.1 wt%). Time of first measurement points were adjusted to t'=0 as the relative time following lag phase. Measurements were based on 40 and 10 tracked biofilm segments for the calibrated OD and CLSM measurements, respectively.

## ARTICLE

**Ionic strength and biofilm flow velocity**

Ionic strength plays a role in regulating the electrostatic interaction between EPS components with implication to environmental biofilms and potential strategies to disrupt pathogenic biofilms.[28,12] The effect is different depending on the type of bacterial strain. For example, for *Pseudomonas aeruginosa*, the biofilm production and motility have been shown to depends on strain and long-term exposure to NaCl.[29] In regards to the mechanical properties, it has been shown that exposure to different mono- and divalent ions could reduce biofilm elasticity and trivalent ions could enhance elasticity, whereas less is known about the role of simple salts on viscosity.[12,13,30] Figure 4 shows the evolving $\tilde{v}_{net}$ values for tracks following inoculation and growth of *Pseudomonas sp.* biofilms under modified nutrient streams with [NaCl] of 0.05 wt%, 0.1 wt% and 0.2 wt%. In all cases $\tilde{v}_{net}$ gradually increased immediately after being visualized until it reached a maximum value. This value and the time required to reach it decreased with increasing NaCl concentration (Figure 4b). The results from a fourth solution with 0 wt% NaCl, showed similar behaviour to the 0.05 wt% sample but did show a reduction in $\tilde{v}_{net}$ in the time interval for these experiments. We note that even for the highly mobile biofilm segments grown under the 0.05 wt% NaCl, the maximum velocity was about 2 orders of magnitude slower than the calculated nutrient flow velocity (ca. 170 mm·h$^{-1}$) based on the nutrient flow rate and the channel dimensions. This confirms that moving biofilm segments were not free-floating, but attached to the microchannel wall.

**Ionic strength and biofilm viscosity**

An equation based on a two-phase flow model, was used to predict $\tilde{v}_{net}$ of the moving biofilm segments.[11]

$$\tilde{v}_{net} = \frac{h(-h^2 + h^2 m + 2hm - 3m)}{m(-9h^2 + 6h^2 m - 4hm + 4h + h^4 m - h^4 + 6h^3 - 4h^3 m + m)} \cdot v_0 \quad (4)$$

Where $h$ is determined from of the tracked biofilms using customized calibration equations similar to (3), $v_0$=0.25 m·h$^{-1}$ is the nutrient stream flow velocity based on the dimensions of the microchannel and the flow rate from the syringe pump, $m$ is the viscosity ratio $\mu_{biofilm}/\mu_{nutrient}$, with the viscosity of the nutrient, $\mu_{nutrient}$, stream having been measured to be 0.96 mPa·s throughout the experiment for all nutrient solutions used here. Solving for time-varying $\mu_{biofilm}$ was accomplished with a spreadsheet with the relevant formulae embedded (Supporting Information). The results are presented in Figure 5. With the exception of the first few hours after biofilm formation, the $\mu_{biofilm}$ was lower with reduced [NaCl]. The lowest values observed of nearly 0.1 Pa·s is lower than any reported values that we are aware of. Biofilms exposed to nutrient solutions containing [NaCl] of 0.05 wt% and 0.1 wt% had pre-thickening $\mu_{biofilm}$ values that were close to constant, whereas biofilm segments grown under 0.2 wt% thickened in the first 3 hours after the end of the lag phase, followed by a gradual increase for the next 20 hours until sudden rapid thickening phase occurred. In all cases, the rapid thickening phase drastically increased $\mu_{biofilm}$ by 1-2 orders of magnitude over a duration of only 5-10 hours. As noted in Fig. 4b, the time at which the rapid thickening occurred scaled inversely with [NaCl]. Due to the sudden nature of the thickening, it appears that the mechanism is not related to a continuous process, such as a build-up of metabolism products or a salting out effect, but the results of a phenotypic change in the bacteria. In addition, there appears to be no relation to the optical density growth curves or calculated height (Supporting Information).

Finally, we use the presented method to revisit a previously reported data set wherein a novel streamer formation mechanism was reported called "sudden partial detachment". Here we verify and add precision to the hypothesis that streamers formed after a transition to high viscosity.[11] Utilizing the calibration method to estimate $h$ discussed above, the results of the time-varying changes to $\mu_{biofilm}$ were obtained (Figure 6). A striking similarity to the thickening phase was observed in Figure 5. In fact, for the same growth conditions ([NaCl]=0.1 wt%), the rapid increase to $\mu_{biofilm}$ was nearly the same, from nearly 6 Pa·s to 300 Pa·s. As well, we can now say that the time of streamer formation observed previously, occurred approximately ten hours after the rapid thickening process finished. Close analysis of the time-lapse video of biofilm growth for [NaCl]=0.1 wt% collected in this study shows similar instances of streamer formation after the rapid thickening occurred. Interestingly, biofilms grown at 0.05 wt% and 0.2 wt% did not show streamer formation after thickening.

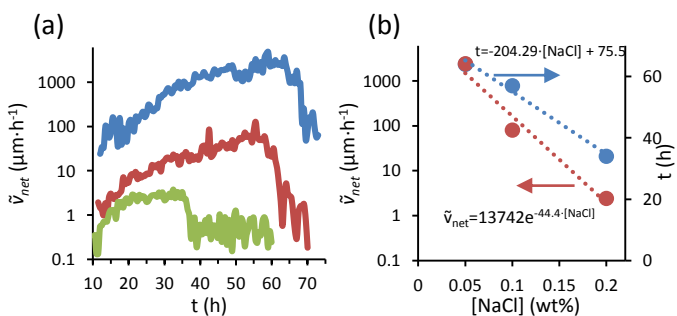

**Figure 4.** (a) Time-dependent $\tilde{v}_{net}$ of tracks for *Pseudomonas sp.* biofilms grown under modified LB nutrient solution containing NaCl concentrations of 0.05 wt% (8.5 mM) in blue, 0.1 wt% (17 M) in red and 0.2 wt% (34 mM) in green. (b) Trends in maximum $\tilde{v}_{net}$ (red) and time to reach maximum $\tilde{v}_{net}$ (blue) versus NaCl concentration and relevant trend lines (dashed) with R$^2$ values that were 0.988 and 0.967, respectively.

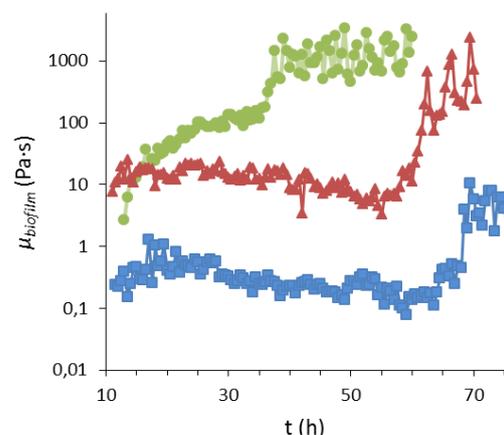

**Figure 5.** Time-dependent $\mu_{biofilm}$ for Pseudomonas sp. biofilms grown under modified LB nutrient solution containing [NaCl] of 0.05 wt% (blue), 0.1 wt% (red) and 0.2 wt% (green).

## Discussion

The present method differentiates itself from other works by enabling continuous measurements of viscosity of intact biofilms, under low, unchanging flow rates. The parallel measurement channels allows for multiple growth environments to be evaluated simultaneously for samples prepared from the same batch, thereby eliminating differences in initial bacterial activity from preculture to preculture. The potential for impact is realized by a comparison to other literature reports. For example, the generally low viscosities measured here are likely the result of low constant flow conditions, which eliminate, or at least strongly reduce, non-Newtonian shear-thickening response.[8] In addition, the very low values measured immediately after the lag phase would not have been observed without the ability to make time-resolved measurements. Reported viscosities for *Pseudomonas* biofilms show a dramatic variation, from tens of Pa·s to $10^5$ Pa·s, and can be as high as $10^8$ Pa·s in others types.[8,31,32] It has not been clear how much of these discrepancies are related to artifacts in the measurement technique, environmental conditions or the type of biofilm. The method here shows that using microfluidics to eliminate shear-response as a potential artifact, studies can focus on specific controllable environmental factors, which up until have not received as much attention. For example in this study, considering relatively small changes to biofilm age and [NaCl], we found that viscosity varied by approximately 4 orders of magnitude. Future studies can undertake systematic studies into the interplay between biofilm viscosity and other environmental factors which may play a role, such as pH, presence and concentration of other salts or chemical species, temperature, shear stresses, solvent polarity, and their combinations.[8,13,33,34] Combined with relevant techniques for complementary, multi-modal *in situ* sensing (e.g. pH, temperature, electrochemical, vibrational spectroscopy) and the high-throughput study of multi-component chemistry in microchannels, the approach shown here also opens the door to combinatorial studies.[35-42]

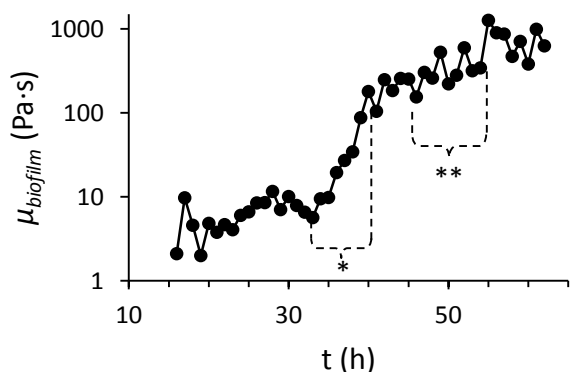

**Figure 6.** A plot of time-dependent $\mu_{biofilm}$ for *Pseudomonas sp.* grown under modified LB nutrient solution containing [NaCl] of 0.1 wt% from previous work previous work.[10] The symbols * and ** show the intervals related to rapid thickening and the formation of streamers, respectively.

## Conclusions

A passive, semi-analytical method is presented for measuring biofilm viscosity using standard microscopy in microchannels. Applied to *Pseudomonas sp.* biofilms, results from time-varying measurements showed strong changes in viscosity within three days after inoculation. The majority of these changes occurred abruptly within a 5-10 hour window. High throughput measurements were facilitated by parallel flow cells on the same microfluidic chip and a scanning microscope stage. This was used to probe the effect of NaCl concentrations in the range 0 to 34 mM (0.2 wt%). In these low concentrations of NaCl, there was a striking difference in viscosities during the measured times. NaCl also appeared to have a strong correlation with the onset time of the detected rapid thickening phase.

## Acknowledgements

The authors wish to thank Marthinus (Otini) Kroukamp for fruitful discussions and the following funding sources, Natural Sciences and Engineering Research Council of Canada, Canada Foundation for Innovation, and Fonds de recherche du Quebec—Nature et technologies for funding support.

## Notes and references

# A microfluidic method for continuous, non-intrusive viscosity measurements of intact biofilms under different chemical conditions

F. Paquet-Mercier,[a] M. Parvinzadeh Gashti,[a] J. Bellavance,[a] S.M. Taghavi,[b] J. Greener[a,*]

## 1. Parallel microfluidic biofilm flow culture device

A schematic of a single flow cell in the microfluidic device used for this work is shown in S1a below. Each microfluidic device six such flow cells, each containing an inlet, outlet and a glass microscope coverslip. Method of fabrication, dimensions and other relevant information are given in the main paper. Figure S1b shows the microfluidic device interfaced with a microscope and automatic computer-controlled stage for accurate placement of the field of view in each channel. The system did not have auto-focus, so care had to be taken to ensure that the device was well-placed on the stage and that all channels were in the same focal plane.

(a) (b)

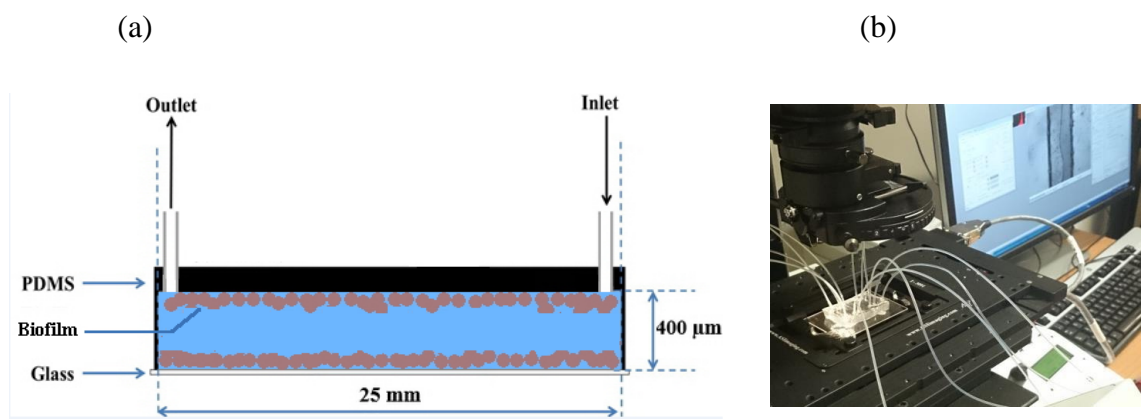

**Figure S1.** (a) A six-channel microfluidic flow cell connected to syringe pump, is mounted on an automatic microscope stage. (b) Schematic of the microfluidic flow cell.

## 2. Velocity measurements in different locations of the microchannel

Figure S2 shows a typical velocity versus time profile for moving biofilm segments for a separate experiment which was conducted with bacteria from different preculture than experiments shown in the main paper. At 8 hours visible biofilm segments began to form in the channel. Shortly after, they began to flow downstream under the shear force of the flowing nutrient solution. In this experiment the $\tilde{v}_{net}$ steadily grew until it reached its maximum value of 64.3 µm·h$^{-1}$ at t=34 h. A shoulder was also noted in the $\tilde{v}_{net}$ vs. time curve. A fitting algorithm using a 2 Gaussian composite peak, found a broad peak centred at t=31 h and a second taller and sharper peak centred at t=34 h. This was attributed to two different behaviours, near and far from the vertical side-walls (i.e., in the microchannel corners), due to strongly varying shear stresses. As this wall effect should be localized to approximately 100 µm from the vertical side wall, we largely avoided this problem by tracking moving biofilm segments further away from the wall. Numerical simulations were conducted using COMSOL multiphysics (v4.2) using physics for "Laminar Flow" and "Transport of Dilute Species" and free tetrahedral meshing with "fine" resolution. Based on these results, we relegated measurements of $v_{net}$ to the centre portion of the channel to avoid wall-induced biases in measurements.

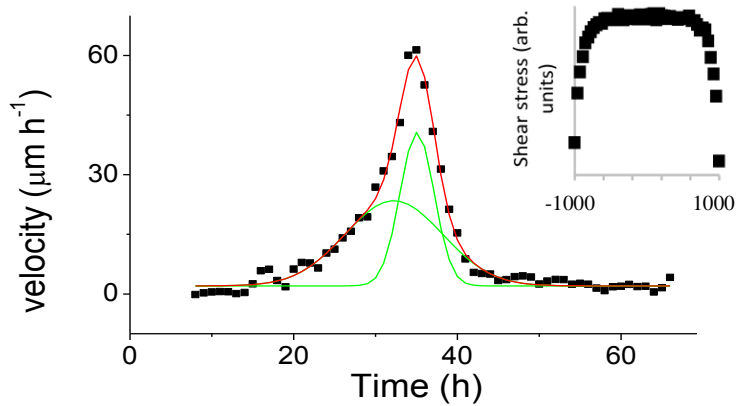

**Figure S2.** Time-dependent ṽ$_{net}$ of tracks (black squares) obtained from a biofilm growing under nutrient solution containing 0.1 wt% NaCl. The red curve is the fitting results using a 2 Gaussian peak model, with the de-convoluted peaks in green. Each point represents the average from 40 different tracks. The inset profiles the change in applied shear stress for different positions along the channel cross-section.

## 3. Growth kinetics

Figure S3 shows the measured optical density as a function of time for biofilms grown under three different NaCl concentrations. These are from the same experiments discussed in the main paper. The semi-log plot shows that the exponential growth region for each biofilm occurs before 20 hours. In all cases, the mature phase, where the OD does not increase further, is after 40 hours.

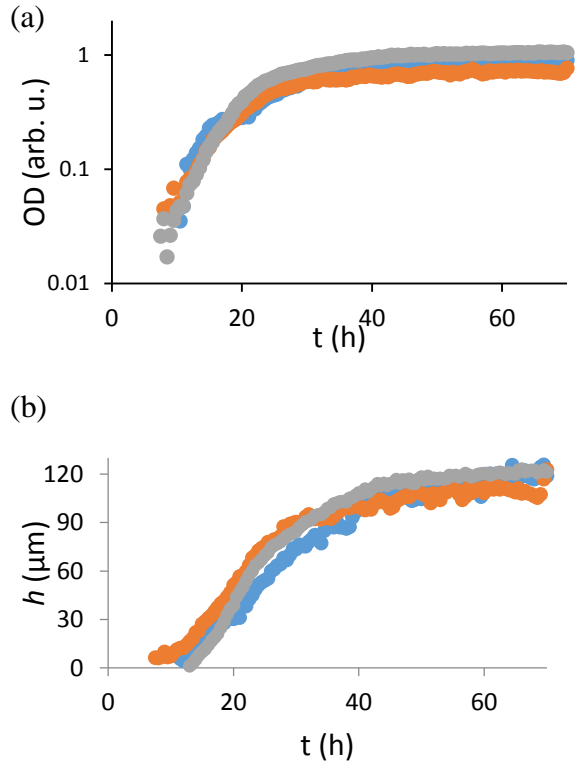

**Figure S3.** (a) Semi-log plot of time-dependent changes to OD and (b) plot of time-dependent *h* for biofilms grown under modified LB nutrient solutions with NaCl concentrations 0.05 wt% (blue), 0.1 wt% (red), 0.2 wt% (green).

## 4. CLSM height measurements

In order to confirm the biofilm height determined from OD measurements in standard microscopy, CLSM was used. The 3D confocal stacks were analyzed using the Fiji bundle for ImageJ. For each time frame, the intensity for each slice of the confocal stack was measured on a region covering the structures that were identified as moving during growth. These measurements resulted in multiple curves as illustrated in figure S4. From these curves the biofilm height was measured by taking the width of the resulting curve as illustrated by the dashed line in figure S4. As the early biofilm is mostly composed of microcolonies made of multiple layers of bacteria, the boundary is harder to define than for biofilm in later growth phases. Coupled with limitations in z-resolution, this is likely the reason that biofilm height is overestimated for the first time points in figure 3 in the main paper.

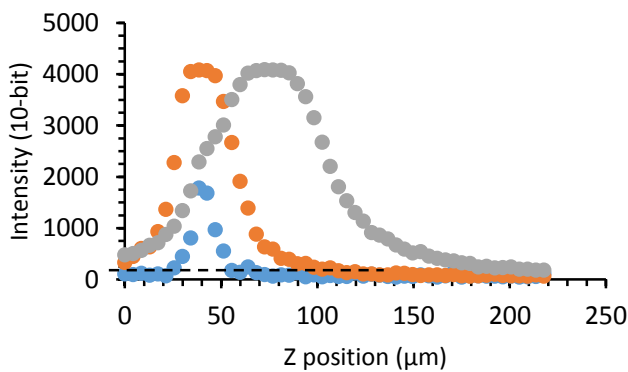

**Figure S4**: Intensity for different Z position for biofilm after 12, 24.5 and 39.5 hours (blue, orange and grey, respectively). Biofilm height was measured by taking the width of the curves.

## 5. *Passivity of CLSM measurements:*

In order to verify that no photoinduced stress was being accumulated by the biofilm, the field of view from a 40x objective was imaged within the larger field of view from a 10x objective. We observed in the confocal stack acquisitioned with the 10x objective inside and outside of the 40x field of view are identical in both biofilm behavior (growth and movement) and intensity. Therefore, there was no noticeable photobleaching of GFP or photodamage to bacteria.